\documentclass[showpacs,preprintnumbers,
superscriptaddress,amsmath,floatfix]{revtex4}

\usepackage{graphicx}
\usepackage{bm} 

\usepackage{epsfig}

\newcommand{\non}{\nonumber\\}

\newcommand{\Diracslash}[1]{#1\llap{/\kern1pt}}
\newcommand{\diracslash}[1]{#1\llap{/\kern0pt}}

\newcommand{\be}{\begin{equation}}
\newcommand{\ee}{\end{equation}}
\newcommand{\bea}{\begin{eqnarray}}
\newcommand{\eea}{\end{eqnarray}}
\newcommand{\ba}[1]{\begin{array}{#1}}
\newcommand{\ea}{\end{array}}

\newcommand{\vk}{\mathbf{k}}

\newcommand{\vg}{\bm{\gamma}}

\newcommand{\GIES}[3]{\mbox{\raisebox{#3} 
{\epsfig{file=#1,scale=#2,clip=true}}~}}

\begin{document}

\title{
A massive high density effective theory
}

\author{Philipp T.\ Reuter}
\email{reuter@triumf.ca}
\affiliation{
TRIUMF, 4004 Wesbrook Mall, Vancouver, BC, Canada, V6T 2A3
}

\date{\today} 

\begin{abstract}
We derive an effective theory for dense, cold and massive quark matter.
To this end, we employ a general effective action formalism where antiquarks
and quarks far from the Fermi surface, as well as hard gluons, are integrated out explicitly.
We show that the 
resulting effective action depends crucially on the projectors used to separate
quarks from antiquarks. If one neglects the quark masses in these projectors, the Feynman rules
of the effective theory involve quark mass insertions which connect quark with antiquark propagators.
Including the quark masses into these projectors, mass insertions do not appear and
the Feynman rules are identical to those found in the zero-mass limit. 
\end{abstract}


\maketitle

\section{Introduction}

Effective theories for quark matter under extreme conditions are of broad interest and of fundamental importance. 
At high temperatures, the hard thermal loop effective theory (HTL) is widely used to study the low-energy 
excitations and the thermodynamics of the hot quark-gluon plasma \cite{braatenpisarski,LeBellac,Blaizot,Kraemmer,Ipp}.
Its counterpart for large quark-chemical potentials and low temperatures is the hard dense loop effective
theory (HDL) \cite{Manuel,Gerhold}.  For temperatures low enough for color superconductivity, high density effective theories
(HDET) have been developed  \cite{hong,hong2,HLSLHDET,schaferefftheory,others,verySchafer,Reuter}.
In \cite{Reuter} an effective framework has been developed which
constitutes the common basis for all these effective theories. In this framework, each effective theory
corresponds to a specific choice of projection operators which are used to separate high from low-energy
quark and gluon modes. Since the details of the high-energy modes are not important for the physics
of interest, these modes are integrated out. Only the low energy modes are kept as actual degrees of freedom
and treated in a self-consistent way.

In the case of the HTL and HDL effective theories, these projectors are defined in momentum space.
For the HTL effective theory, e.g., the irrelevant quark and gluon modes have hard momenta of the order of the
temperature $T$ or larger. Relevant modes carry soft momenta of the order of $gT$,
where the gauge coupling $g$ is small, $g\ll 1$, for large temperatures, $T\gg \Lambda_{\rm QCD}$.
In the case of the HDETs, quarks close to the Fermi surface, $|k-k_F|<\Lambda_{\rm q}$,
and soft gluons with $p<\Lambda_{\rm gl}$ are included, whereas hard gluons, $p>\Lambda_{\rm gl}$,
quarks far from the Fermi surface, $|k-k_F|>\Lambda_{\rm q}$, and antiquarks
 are integrated out. Due to the special kinematics of 
quarks scattering along the Fermi surface and the dominance of
long-range, almost static magnetic gluons, different cutoffs for quarks and gluons have been considered
 $\Lambda_{\rm q} \ll \Lambda_{\rm gl}$ \cite{Reuter,Reuter2,Reuter3,NFL}. 
 In \cite{Reuter,Reuter2,Reuter3} the cutoffs $\Lambda_{\rm q} \sim g k_F$ and $\Lambda_{\rm gl} \sim k_F$
 have been considered. Antiquarks, on the other hand,
require projectors with a non-trivial Dirac structure. For that purpose, the projectors
\be\label{L0}
\Lambda^e_{\vk,0} = \frac{1}{2}
\left( 1 + e \gamma_0 \vg \cdot
\hat {\bf k}  \right)\;,
\ee
are used \cite{hong,hong2,HLSLHDET,schaferefftheory,others,verySchafer},
which project on the solutions of the free, massless Dirac equation with positive ($e=+$) or
 negative ($e=-$) energies.
The energy projectors (\ref{L0}) commute with the projectors for
chirality, so that the quark spinors in the HDET can also be labeled with a chiral quantum number.

This chiral HDET is also used for massive quarks \cite{Son,Savage,Schafer,Kryjevski,Kryjevski2,Schaferms}. Then, however,
the quark spinors contain an admixture of antiquarks even without interactions.
Consequently, additional terms must be included into the theory, where mass insertions connect quark 
with antiquark propagators. 
The number of such terms, unfortunately, increases rapidly if one includes higher orders in $m/k_F$,
and the HDET becomes rather difficult to manage. It is therefore a worthwhile goal to find an 
alternative, non-chiral representation where the quark masses do not complicate the Feynman rules.
For that reason, we refrain from adding any terms by hand in this work but rather
continue the venue of \cite{Reuter} and {\it explicitly derive} all terms that include quark masses. This approach allows
to analyze systematically how the Feynman rules depend on the representation 
of the quark spinors in the effective theory. In addition to the projectors (\ref{L0}), we will also consider 
the projectors on the solutions of the free, but {\it massive} Dirac equation \cite{Pisarski}
\be\label{Lm}
\Lambda^e_{\vk,m} = \frac{1}{2 E_k} \, 
\left[ E_k + e \gamma_0 \left(\vg \cdot
{\bf k} + m \right) \right]\;,
\ee 
where 
\be\label{Ek}
E_k = \sqrt{{\bf k}^2 +m^2}
\ee
is the relativistic single-particle energy. The projectors (\ref{Lm}) differ from the more commonly used massive 
projectors \cite{Gross} by a $\gamma_0$ and the normalization, but have the advantage that in the limit
$m\rightarrow 0$ they remain regular and converge to the massless projectors (\ref{L0}), 
$\Lambda^e_{\vk,m} \rightarrow \Lambda^e_{\vk,0}$.  For $m>0$ the projectors (\ref{Lm}) do 
not commute with the projectors for chirality. The quark spinors in the effective theory are therefore no longer
eigenstates of chirality. Moreover, quark and antiquark states are not mixed anymore in the limit of no interactions,
so that the antiquark fields can be integrated out completely. This crucially effects
the Feynman rules of the resulting effective theory. In order to discriminate this non-chiral theory from the chiral HDET,
we will refer to it as the {\it massive} high density effective theory (mHDET). 
We show that in the mHDET one has the same Feynman rules as in the zero-mass limit. Mass insertions are absent
and antiquark propagators can only appear after the emission or absorption of a gluon. The quark masses enter
exclusively through the quark energy (\ref{Ek}).  This significantly reduces the
number of diagrams as compared to the HDET. In order to calculate mass corrections in the mHDET,
one considers the Taylor expansion
\bea\label{taylor}
E_k= k+m^2/2k-m^4/8k^3+\cdots
\eea
in the quark and antiquark propagators of any given term. We hope that in future work, this will help to maintain the overview when organizing the contributions to a given order in $m/k_F$. Furthermore, we expect the actual calculations become more transparent,
because mass corrections originate only from the expansion (\ref{taylor}) and no additional diagrams are required.

This paper is organized as follows. In Sec.\ \ref{formalism} the general effective action formalism of \cite{Reuter}
is briefly reviewed. In Sec.\ \ref{mHDET} the formalism is used to derive the non-chiral mHDET, i.e.\ using 
the massive projectors (\ref{Lm}). In Sec.\ \ref{HDET} the massless projectors (\ref{Lm}) are used to derive the 
chiral HDET. In both the mHDET and the HDET mass corrections are contained to all orders. 
In the HDET these occur as mass insertions in the Feynman diagrams, whereas in the mHDET the quark masses are 
included in the relativistic quark energies $E_k$.
Sec.\ \ref{discus} the results are summarized.
Our units are $\hbar=c=k_B=1$. 4-vectors are denoted by
capital letters, $K^\mu = (k_0, {\bf k})$, with ${\bf k}$ being a
3-vector of modulus $|{\bf k}| \equiv k$ and direction
$\hat{\bf k}\equiv {\bf k}/k$. 
We work in compact Euclidean space-time with 
volume $V/T$, where $V$
is the 3-volume and $T$ the temperature of the system. 
Since space-time is compact, energy-momentum space is
discretized, with sums $(T/V)\sum_{K} \equiv T\sum_n (1/V) \sum_{\bf
k}$. For a large 3-volume $V$, the sum over 3-momenta
can be approximated by an integral, $(1/V)\sum_{\bf k} \simeq
\int d^3 {\bf k}/(2 \pi)^3$.

\section{The Effective action for cold, dense and massive quarks}
\subsection{The general effective action formalism}
\label{formalism}

We review the general effective action formalism developed in \cite{Reuter}.  In the Sec.\ \ref{mHDET} and \ref{HDET}
it will be used to derive the mHDET and the HDET.  In the Nambu-Gor'kov basis \cite{DHRreview} the quark spinors are
\be
\Psi \equiv \left( \begin{array}{c}
                    \psi \\
                    \psi_C \end{array} \right) \;\; , \;\;\;\;
\bar{\Psi} \equiv ( \bar{\psi} , \bar{\psi}_C )\,\, ,
\ee 
where $\psi_C\equiv \bar \psi ^\dagger$ is the charge-conjugate spinor and $C \equiv i \gamma^2 \gamma_0$
is the charge-conjugation matrix. This representation is advantageous for applications to color superconductivity.
All arguments of the present work, however, can be understood without considering the charge-conjugate sector.
The quark fields are split up into relevant and irrelevant modes according to 
\be \label{project}
\Psi_1 \equiv {\cal P}_1 \, \Psi\;\; , \;\;\;\;
\Psi_2 \equiv {\cal P}_2 \, \Psi \;\; , \;\;\;\;
\bar{\Psi}_1 \equiv \bar{\Psi} \, \gamma_0 {\cal P}_1 \gamma_0 
\;\; , \;\;\;\;
\bar{\Psi}_2 \equiv \bar{\Psi} \, \gamma_0 {\cal P}_2 \gamma_0 
\;\;,
\ee
where ${\cal P}_1$ and ${\cal P}_2\equiv 1- {\cal P}_1$ are projection operators.  They will be specified
in Secs.\ \ref{mHDET} and \ref{HDET}.
With that the quark partition function can be written as
\be \label{Zq3}
{\cal Z}_q[A] = \int \prod_{\ell=1,2} {\cal D} \bar{\Psi}_\ell\, 
{\cal D} \Psi_\ell \, \exp \left( \frac{1}{2} \sum_{n,m=1,2} \bar{\Psi}_n \, 
{\cal G}^{-1}_{nm} \, \Psi_m \right)
\,\, ,
\ee
where for the sake of compactness we suppressed all the momentum arguments of the 
fields and the matrices ${\cal G}^{-1}_{nm}$, where $n,m \in \{1,2\}$. The latter are defined as
\bea\label{Gmn}
{\cal G}_{nm}^{-1}(K,Q) \equiv \gamma_0 {\cal P}_n \gamma_0 
\left[{\cal G}^{-1}(K,Q)\right]{\cal P}_m \;,
\eea
where 
\bea
{\cal G}^{-1}(K,Q) \equiv  {\cal G}_{0}^{-1}(K,Q) + g {\cal A}(K,Q)\;.
\eea
The free inverse quark propagator is given by
\be 
\label{G0FT}
{\cal G}_0^{-1}(K,Q) = \frac{1}{T} \left( \begin{array}{cc}
                             [G_0^+]^{-1}(K) & 0 \\
                              0 & [G_0^-]^{-1}(K) \end{array} \right)
\delta^{(4)}_{K,Q} \;, 
\ee
where $[G_0^\pm]^{-1}(K) \equiv \Diracslash{K} \pm \mu \gamma_0 - m
$.
The Grassmann integration over the irrelevant fields $\bar{\Psi}_2,\,
\Psi_2$ can be done exactly, if one redefines them such that
the mixed terms $\sim {\cal G}_{nm}^{-1}$, $n \neq m$, are eliminated.
To this end, substitute 
\be\label{shift}
\Upsilon_2 \equiv \Psi_2 + {\cal G}_{22}\, {\cal G}_{21}^{-1}\, \Psi_1\;\; ,
\;\;\;\; 
\bar{\Upsilon}_2 \equiv \bar{\Psi}_2 + \bar{\Psi}_1\, 
{\cal G}_{12}^{-1}\, {\cal G}_{22}\; .
\ee 
The result is
\be \label{Zq4}
{\cal Z}_q[A] 
= \int {\cal D} \bar{\Psi}_1\, 
{\cal D} \Psi_1 \, \exp \left[ \frac{1}{2} \, \bar{\Psi}_1 
\left( {\cal G}^{-1}_{11} - {\cal G}^{-1}_{12} \, {\cal G}_{22}
\, {\cal G}^{-1}_{21} \right) \Psi_1
+ \frac{1}{2}\, {\rm Tr}_q \ln {\cal G}_{22}^{-1} \right] \,\, .
\ee
The trace runs over 4-momenta and Nambu-Gor'kov, fundamental color, 
flavor, and Dirac indices. We indicate this by the subscript ``$q$''. 
The term ${\cal G}_{22}$ has an expansion in powers of $g$ times the gluon field,
which can be symbolically written as
\be\label{expandG22}
{\cal G}_{22} = \left( 1 +  {\cal G}_{0,22}\,g {\cal A} \right)^{-1}
{\cal G}_{0,22}\;.
\ee

The next step is to integrate out hard gluon modes.  Including the
gluonic action $S_A[A]$, cf.\ eqs.\ (38-40) in \cite{Reuter}, we have
\begin{subequations} \label{ZQCD2}
\bea 
{\cal Z} & = &
\int {\cal D} \bar{\Psi}_1 \,  {\cal D} \Psi_1\, 
{\cal D} A \, \exp \left\{ S[A,\bar{\Psi}_1,\Psi_1] \right\}\;, \\
S[A,\bar{\Psi}_1,\Psi_1] & \equiv &
S_A[A] + \frac{1}{2} \, \bar{\Psi}_1 
\left( {\cal G}_{11}^{-1}-
{\cal G}_{12}^{-1} {\cal G}_{22} {\cal G}_{21}^{-1}\right)\![A]\;\Psi_1
+ \frac{1}{2}\, {\rm Tr}_q \ln {\cal G}_{22}^{-1}[A] \;. \label{S}
\eea
\end{subequations}
For the sake of clarity, we restored
the functional dependence of the quark propagators
${\cal G}_{mn}$ on the gluon field $A$.
Similar to the treatment of fermions
we introduce projectors ${\cal Q}_1,\, {\cal Q}_2$ for
soft and hard gluon modes, respectively,
\be
A_1  \equiv {\cal Q}_1 \, A \;\; , \;\;\;\; A_2 \equiv {\cal Q}_2 \,
A\;,
\ee
where
\begin{subequations} \label{Q12}
\bea
{\cal Q}_1(P_1,P_2) & \equiv & \Theta(\Lambda_{\rm gl} -p_1) \, \delta^{(4)}_{P_1,P_2}
\; , \\
{\cal Q}_2(P_1,P_2) & \equiv & \Theta(p_1-\Lambda_{\rm gl}) \, \delta^{(4)}_{P_1,P_2}
\; .
\eea
\end{subequations}
The gluon cutoff momentum $\Lambda_{\rm gl}$ defining which gluons are soft or hard 
does not have to be specified at this point. Inserting $A \equiv A_1 + A_2$ into 
Eq.\ (\ref{ZQCD2}) the 
action $S[A,\bar{\Psi}_1,\Psi_1]$ can be sorted with respect to
powers of the hard gluon field,
\be \label{expansion}
S[A,\bar{\Psi}_1,\Psi_1] = S[A_1,\bar{\Psi}_1,\Psi_1]
+ A_2 {\cal J} [A_1,\bar{\Psi}_1,\Psi_1] - \frac{1}{2}\,
A_2 \, \Delta^{-1}_{22}[A_1, \bar{\Psi}_1,\Psi_1]\, A_2 
+ S_I[A_1,A_2,\bar{\Psi}_1,\Psi_1] \;.
\ee 
The first term in this expansion, containing no hard gluon fields
at all, is simply the action (\ref{S}), with $A$ replaced by
the relevant gluon field $A_1$.
The second term, $A_2 {\cal J}$,
contains a single power of the hard gluon field, where
\be \label{J}
{\cal J} [A_1,\bar{\Psi}_1,\Psi_1] \equiv 
\left. \frac{\delta S[A, \bar{\Psi}_1,\Psi_1]}{\delta A_2}
\right|_{A_2 =0} \,\, ,
\ee
defines the hard gluon "current". For cold and dense quark matter it can be shown \cite{Reuter}
that due to the conservation of energy and momentum the only contributions to $\cal J$ come from ${\cal G}_{11}^{-1}-
{\cal G}_{12}^{-1} {\cal G}_{22} {\cal G}_{21}^{-1}$, where a quark is scattered 
along the the Fermi surface by absorbing or emitting one or more hard gluons.
The term quadratic in $A_2$ is defined as
\be
\Delta_{22}^{-1}[A_1,\bar{\Psi}_1,\Psi_1]\equiv 
- \left. \frac{\delta^2 S[A, \bar{\Psi}_1,\Psi_1]}{\delta A_2\,\delta A_2}
\right|_{A_2 =0} \equiv \Delta_{0,22}^{-1} + 
\Pi_{22}[A_1, \bar{\Psi}_1, \Psi_1]\,\, .
\ee
Here, $\Delta_{0,22}^{-1}$ is the free inverse propagator 
for hard gluons. 
The ``self-energy'' of the hard gluons $\Pi_{22}$ is comprised of
contributions from ${\cal G}_{11}^{-1}-
{\cal G}_{12}^{-1} {\cal G}_{22} {\cal G}_{21}^{-1}$, from the 3- and 4-gluon vertex 
contained in $S_A$, and from the quark loops ${\rm Tr}_q \ln {\cal G}_{22}^{-1}$
with exactly two hard gluon legs. Note that for 
$\Lambda_{\rm gl}\sim k_F$, the latter are of the order of $g^2k_F^2$
 and therefore are small as compared to $\Delta_{0,22}^{-1}$ which is of order $k_F^2$.
Furthermore, note that for small temperatures, $T \ll \mu$, contributions from gluon 
and Fadeev-Popov ghost loops are negligible as compared to those from the quark loops.
The term $S_I$, finally, contains all terms of higher order
in $A_2$. In \cite{Reuter} it is shown that those are not needed to reproduce the HTL, HDL 
effective theories or the HDET. Furthermore, they contribute only beyond subleading order
 to the color superconducting gap parameter \cite{Reuter}. Therefore we will also neglect $S_I$ 
in this work. Its restoration can be done straightforwardly.

With these approximations made the partition function reads
\be \label{Z4}
{\cal Z} =  \int {\cal D} \bar{\Psi}_1 \, {\cal D} \Psi_1 {\cal D} A_1\,
\exp\{S_{\rm eff} [A_1,\bar{\Psi}_1, \Psi_1 ] \}\;,
\ee
where the effective action is defined as
\bea \label{gefac}
S_{\rm eff} [A_1,\bar{\Psi}_1, \Psi_1 ] & \equiv & 
S_A[A_1] + \frac{1}{2} \, \bar{\Psi}_1 
\left( {\cal G}_{11}^{-1}-
{\cal G}_{12}^{-1} {\cal G}_{22} {\cal G}_{21}^{-1}\right)\![A_1]\;\Psi_1
+ \frac{1}{2}\, {\rm Tr}_q \ln {\cal G}_{22}^{-1}[A_1]  \non
&  &
- \frac{1}{2}\,  {\rm Tr}_g \ln \Delta_{22}^{-1}[A_1,\bar{\Psi}_1,\Psi_1] 
+ \; \frac{1}{2} \, {\cal J}[A_1,\bar{\Psi}_1,\Psi_1]   \,
\Delta_{22}[A_1,\bar{\Psi}_1,\Psi_1] 
\, {\cal J}[A_1,\bar{\Psi}_1,\Psi_1]  \; . 
\label{Seff}
\eea
This is the desired action for the effective theory describing the
interaction of relevant quark modes, $\bar{\Psi}_1, \Psi_1$, and
soft gluons, $A_1$. The form of the propagators ${\cal G}_{mn}$,
the hard gluon current $\cal J$, and the 
self-energy for hard gluons, $\Pi_{22}$ determine the Feynman rules of the effective theory.
These have been studied extensively in the limit of massless quarks in \cite{Reuter}.
Non-zero quark masses enter the theory only through the propagators ${\cal G}_{mn}$.
 These appear directly in eq.\ (\ref{gefac}), but also indirectly through $\cal J$ and $\Pi_{22}$.
It is therefore important to analyze how the propagators ${\cal G}_{mn}$ are modified 
by the quark masses and how this depends on the quark projectors ${\cal P}_{1,2}$.
As already indicated in the Introduction, the structure of the propagators ${\cal G}_{mn}$
remains intact if the massive projectors (\ref{Lm}) are used and consequently the 
Feynman rules are the same as in the zero-mass limit. In the case of the massless 
projectors (\ref{L0}) modifications in the propagators ${\cal G}_{mn}$ and, correspondingly,
also in the Feynman rules occur.

\subsection{Quark masses in the mHDET}\label{mHDET}

The projectors for the mHDET are defined as
\begin{subequations} \label{P12m}
\bea
{\cal P}_1(K,Q) & \equiv & \left( \begin{array}{cc}
 \Lambda_{{\bf k},m}^+ & 0 \\
 0 & \Lambda_{{\bf k},m}^- \end{array} \right) \, 
\Theta(\Lambda_{\rm q} - | k - k_F|) \, \delta^{(4)}_{K,Q} \;, \\
{\cal P}_2(K,Q) & \equiv & \left( \begin{array}{cc} 
\Lambda_{{\bf k},m}^- + \Lambda_{{\bf k},m}^+\, \Theta(| k - k_F| - \Lambda_{\rm q})
& 0 \\
0 &  \Lambda_{{\bf k},m}^+ + \Lambda_{{\bf k},m}^-\, \Theta(| k - k_F| -
\Lambda_{\rm q}) \end{array} \right)\, 
\delta^{(4)}_{K,Q}\;,
\eea
\end{subequations}
where $\Lambda_{{\bf k},m}^e$ are given in eq.\ (\ref{Lm}). Hence, ${\cal P}_1$
projects onto quarks close to the Fermi surface and ${\cal P}_2$ onto 
quarks far from the Fermi surface and antiquarks. Note that,
for the Nambu-Gor'kov components corresponding to charge-conjugate
particles, the role of the projectors onto positive and negative energy
states is reversed with respect to the Nambu-Gor'kov components
corresponding to particles.
The reason is that, loosely speaking, a particle is actually a 
charge-conjugate antiparticle. 

With the projectors (\ref{P12m}) one can expand
\bea\label{G0m}
 [G_0^{\pm}]^{-1}(K)=
\sum_{e=\pm}(k_0\pm\mu -eE_\vk)\gamma_0\,\Lambda_{\vk,m}^{e}
\eea
and hence
\begin{subequations} \label{G0mn}
\bea 
{\cal G}_{0,11}^{-1}(K,Q) &=& \frac{1}{T} \left( \begin{array}{cc}
                             (k_0+\mu-E_k)\gamma_0\,\Lambda_{\vk,m}^{+} & 0 \\
                              0 & (k_0-\mu+E_k)\gamma_0\,\Lambda_{\vk,m}^{-} \end{array} \right)
 \Theta(\Lambda_{\rm q} - | k - k_F|)\;\delta^{(4)}_{K,Q} \,\,, \\  \label{G011}
{\cal G}_{0,22}^{-1}(K,Q) &=& \frac{1}{T} \left( \begin{array}{cc}
                             (k_0+\mu-E_k)\gamma_0 \,\Lambda_{\vk,m}^{+}& 0 \\
                              0 & (k_0-\mu+E_k)\gamma_0\,\Lambda_{\vk,m}^{-} \end{array} \right)
 \Theta( | k - k_F|-\Lambda_{\rm q})\;\delta^{(4)}_{K,Q}  \nonumber \\  
&+&				\frac{1}{T} \left( \begin{array}{cc}
                              	(k_0+\mu+E_k)\gamma_0\,\Lambda_{\vk,m}^{-} & 0 \\
                              0 & (k_0-\mu-E_k)\gamma_0\,\Lambda_{\vk,m}^{+} \end{array} \right)
\delta^{(4)}_{K,Q} \,\,, \\ \label{G022}
{\cal G}_{0,12}^{-1}(K,Q) &=&{\cal G}_{0,21}^{-1}(K,Q)= 0 \label{G012}\;.
\eea
\end{subequations}
It follows that the quark masses enter the propagators for relevant and 
irrelevant quark modes, ${\cal G}_{0,11}$ and ${\cal G}_{0,22}$ through the energy $E_k$.
This is the only modification as compared to the massless case discussed in \cite{Reuter}. 
Consequently, in the basis of massive projectors the effective action 
for dense quark matter has the same form as in the zero-mass limit and all Feynman rules
remain unchanged. The transition from the zero-mass limit to non-zero quark
masses is therefore accomplished by substituting
\bea
k_0\pm\mu-ek \; \longrightarrow \; k_0\pm\mu-eE_k
\eea
in all quark propagators in the effective action (49) in \cite{Reuter}. From eq.\ (\ref{G012})
we conclude that there is no coupling of quarks and antiquarks through quark masses.

\subsection{Quark masses in the HDET}\label{HDET}

\begin{figure}[ht]
\includegraphics[width=8cm]{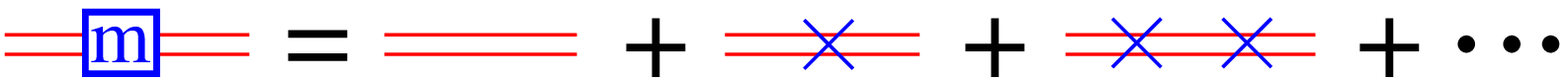}
\caption{The free propagator for irrelevant quarks in the chiral basis 
being resummed with mass insertions. The free irrelevant
quark propagators ${\cal G}_{0,22}$ are denoted by (red) double lines, 
the mass insertions by (blue) crosses.}
\label{MG022}
\end{figure}

\begin{figure}[ht]
\includegraphics[width=8cm]{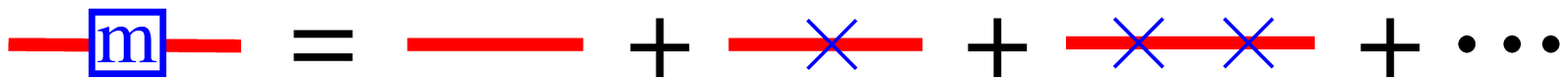}
\caption{The propagator ${\cal G}_{22}$, cf.\ eq.\ (\ref{expandG22}), 
for irrelevant quarks in the chiral basis 
being resummed with mass insertions.}
\label{MG22}
\end{figure}

For the HDET the projectors ${\cal P}_{1,2}$ are defined as in eq.\ (\ref{P12m}) with the massive energy projectors 
$\Lambda^e_{\vk,m}$ being replaced by the massless $\Lambda^e_{\vk,0}$. With that one has
\be\label{G0rep1}
[G_0^\pm]^{-1}(K) =\sum\limits_{e=\pm}(k_0\pm\mu -ek)\gamma_0\,\Lambda_{\vk,0}^{e}-m\;,
\ee
where in contrast to eq.\ (\ref{G0m}) the Dirac structure of the mass term differs by a factor of $\gamma_0$ 
from the remaining parts of $[G_0^\pm]^{-1}$. It follows that
\begin{subequations}
\bea 
 \label{G0mn2}
{\cal G}_{0,11}^{-1}(K,Q) &=& \frac{1}{T} \left( \begin{array}{cc}
                             (k_0+\mu-k)\gamma_0\Lambda^+_{\vk,0} & 0 \\
                              0 & (k_0-\mu+k)\gamma_0 \Lambda^-_{\vk,0}\end{array} \right)
 \Theta(\Lambda_{\rm q} - | k - k_F|)\;\delta^{(4)}_{K,Q} \,\,,   \label{invG011} \\
{\cal G}_{0,22}^{-1}(K,Q) &=& \frac{1}{T} \left( \begin{array}{cc}
                             (k_0+\mu-k)\gamma_0\Lambda^+_{\vk,0} & 0 \\
                              0 & (k_0-\mu+k)\gamma_0\Lambda^-_{\vk,0} \end{array} \right)
 \Theta( | k - k_F|-\Lambda_{\rm q})\;\delta^{(4)}_{K,Q}  \nonumber \\  
&+&				\frac{1}{T} \left( \begin{array}{cc}
                              	(k_0+\mu+k)\gamma_0\Lambda^-_{\vk,0} & 0 \\
                              0 & (k_0-\mu-k)\gamma_0\Lambda^+_{\vk,0} \end{array} \right)
\delta^{(4)}_{K,Q}
\nonumber \\  
&-&				\frac{m}{T} 
\;\Theta( | k - k_F|-\Lambda_{\rm q})\;\delta^{(4)}_{K,Q} \,\,, \label{invG022}  \\
{\cal G}_{0,12}^{-1}(K,Q) &=&{\cal G}_{0,21}^{-1}(K,Q)= -\frac{m}{T}\,
 \Theta(\Lambda_{\rm q} - | k - k_F|)\;\delta^{(4)}_{K,Q}\;. \label{invG012}
\eea
\end{subequations}
Thus, in the HDET the quark masses occur in ${\cal G}_{0,12}^{-1}$ and ${\cal G}_{0,21}^{-1}$,
as well as in the free inverse propagator of irrelevant quark modes, ${\cal G}_{0,22}^{-1}$. 
Inverting the latter, one finds that the propagators of the quarks far from the Fermi surface and antiquarks
are connected by mass insertions. For the upper left component of ${\cal G}_{0,22}$, e.g., one finds
\bea\label{G022M}
\left[\frac{ \Theta( | k - k_F|-\Lambda_{\rm q})\Lambda^+_{\vk,0}\gamma_0}{k_0+\mu-k}
+\frac{\Lambda^-_{\vk,0}\gamma_0}{k_0+\mu+k}\right]
+m\,\frac{ \Theta( | k - k_F|-\Lambda_{\rm q})}{(k_0+\mu)^2-k^2}
+\left[\frac{m\;\Theta(\Lambda_{\rm q} - | k - k_F|)}{(k_0+\mu)^2-k^2}\right]^2
\sum_{e=\pm}\frac{\Lambda^e_{\vk,0}\gamma_0}{k_0+\mu-ek}+\cdots\;.
\eea
Diagrammatically, we represent this resummation with quark masses by a $m$ in a box,
cf.\ Fig \ref{MG022}.
Each cross stands for a mass insertion. 
Also the propagator ${\cal G}_{22}$, cf.\ eq.\ (\ref{expandG22}), is resummed
by mass insertions, which we denote analogously, cf.\ Fig.\ \ref{MG22}.

\begin{figure}[ht]
\includegraphics[width=12cm]{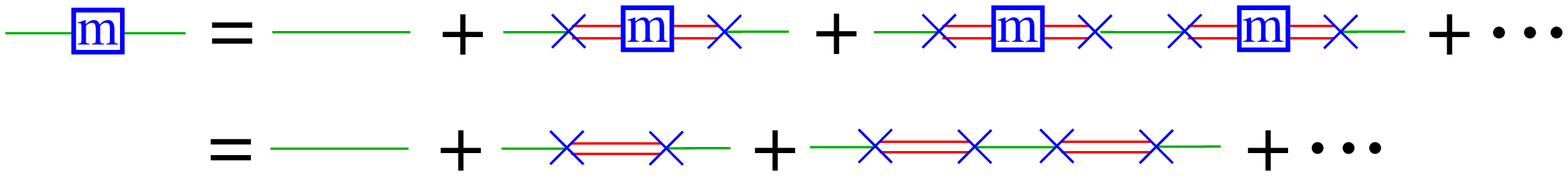}
\caption{The free propagator for relevant quarks in the chiral basis 
being resummed with mass insertions. The free irrelevant
quark propagators ${\cal G}_{0,22}$ are denoted by (red) double lines, 
the mass insertions by (blue) crosses.}
\label{MG11}
\end{figure}

\noindent
In order to fascilitate the interpretation of the Feynman rules, we introduce
\be\label{1212}
{\cal G}^{-1}_{11} - {\cal G}^{-1}_{12} \, {\cal G}_{22}
\, {\cal G}^{-1}_{21} 
\equiv {\cal G}_{0,11}^{-1}  + g {\cal B}+{\cal M}\;,
\ee
where 
\begin{subequations}
\bea \label{B}
g{\cal B} &\equiv& g {\cal A}_{11} - g {\cal A}_{12} \, 
{\cal G}_{22} \, g {\cal A}_{21}\;,\\
\label{M}
{\cal M}&\equiv& \frac{m}{T}\,{\cal G}_{22}\,g{\cal A}+g{\cal A}\,{\cal G}_{22}\,\frac{m}{T} -\frac{m}{T}\,{\cal G}_{22}\,\frac{m}{T}\;,
\eea
\end{subequations}
The term $\cal B$ contains the soft gluon field and its the resummation with internal irrelevant 
quark and antiquark lines, ${\cal G}_{0,22}$. The term ${\cal M}$ is non-zero only for non-vanishing quark masses. 
It resums the coupling of relevant quarks with soft gluons, $\frac{1}{2}\bar \Psi g{\cal B}\Psi$, 
and the free propagator of relevant quarks, ${\cal G}_{0,11}$, with mass insertions.
We denote ${\cal G}_{0,11}$ as in Fig.\ \ref{MG11}.

\begin{figure}[ht]
\includegraphics[width=8cm]{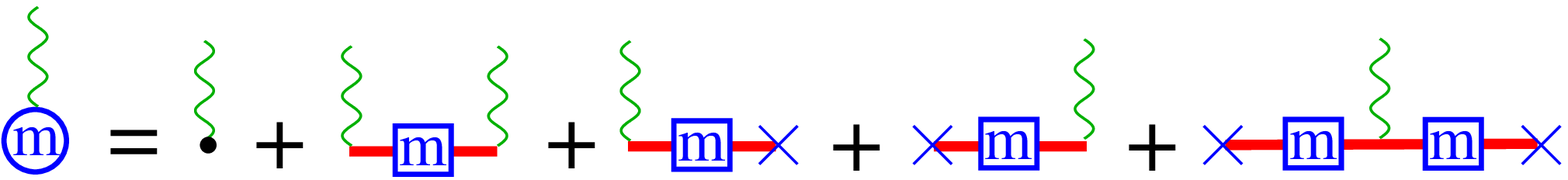}
\caption{The effective coupling of relevant quarks with soft gluons in the chiral basis.
The first two terms correspond to $g\cal B$, cf.\ eq.\ (\ref{B}), the last three terms come from $\cal M$,
 cf.\ eq.\ (\ref{M}).}
\label{Mqg}
\end{figure}

\noindent
Note that all mass insertions in the internal irrelevant quark propagators vanish because
they would require the inclusion of quarks far from the Fermi surface. Since, however,
mass insertions are diagonal in momentum space these modes cannot occur in ${\cal G}_{0,11}$
 due to momentum conservation. Furthermore,
we denote the resummed quark-gluon coupling as in Fig.\ \ref{Mqg}, where the first two terms
come from $g\cal B$ and the last three from $\cal M$.

\noindent
A careful analysis shows that the terms linear in both the gluon field and the quark mass dissappear
in the limit of vanishing gluon momenta \cite{Schafer}.
Otherwise these terms have to be kept and, in particular, must be included in the hard
gluon current
\be \label{J_BM}
{\cal J}_{(g{\cal B}+{\cal M})}[A_1,\bar{\Psi}_1,\Psi_1] 
\equiv \frac{1}{2} \, \bar{\Psi}_1 \, \left[  \frac{\delta (
g{\cal B}+{\cal M})}{\delta A_2}\right]_{A_2=0} \Psi_1
 \;.
\ee 
In our diagrammatic language the hard gluon current reads
\bea\label{JMB}
{\cal J}_{(g{\cal B}+{\cal M})}\Delta_{0,22}=\GIES{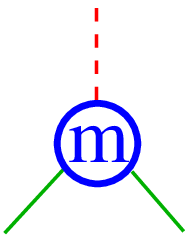}{0.5}{-4.2mm}\;.
\eea
The term $\cal M$ has also to be included into the self-energy of hard gluons.
Again we combine it with $g\cal B$ and obtain
\be\label{Pi_BM}
\Pi_{(g{\cal B}+{\cal M})}[A_1, \bar{\Psi}_1, \Psi_1] \equiv
- \frac{1}{2} \, \bar{\Psi}_1 \, \left[\frac{\delta^2
(g{\cal B}+{\cal M})}{\delta A_2 \delta A_2}\right]_{A_2=0} \Psi_1 \;,
\ee
Diagrammatically this corresponds to
\bea\label{PiMB}
\Delta_{0,22}\Pi_{(g{\cal B}+{\cal M})}\Delta_{0,22}=\GIES{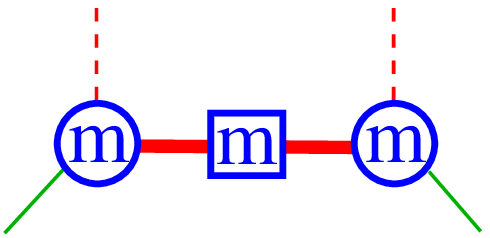}{0.5}{-4.2mm}\;.
\eea
With that all building blocks of the HDET are complete to all orders in $m/k_F$. We found that in the HDET the
quark masses appear as mass insertions in all quark propagators and quark-gluon vertices of the effective
theory, cf.\ Figs.\
 \ref{MG022}-\ref{Mqg} and eqs.\ (\ref{JMB},\ref{PiMB}).
This results in additional diagrams that have to be considered as one considers corrections from quark masses.

\section{Summary}\label{discus}

We showed that the way how quark masses enter effective theories for dense and cold quark matter crucially depends on 
the representation of the quark spinors. We have considered a chiral and a non-chiral representation,
and have analyzed the Feynman rules in both cases. The chiral representation corresponds to the usual HDET where
the quark spinors are solutions of the free, massless Dirac equation. In this effective theory one has to resum all 
quark propagators and all quark-gluon vertices with mass insertions, which results in additional diagrams as compared 
to the limit of zero quark masses. In the non-chiral version the quark spinors 
are solutions of the free, {\it massive} Dirac equation. In the corresponding theory, which we refer to as the mHDET,
the quark masses enter only through the relativistic quark energy, $E_k=\sqrt{k^2+m^2}$. Non-zero quark masses are fully accounted for by replacing $k \rightarrow E_k$ in all quark 
and antiquark propagators. Mass corrections can be obtained systematically by expanding $E_k$, and no additional Feynman 
diagrams are required. These results facilitate the organization mass corrections in dense and cold quark matter to 
at any given order in $m/k_F$ and help to keep the necessary approximations transparent.

\section*{Acknowledgments}

I thank Jens Braun, Michael Forbes, David Kaplan, Rob Pisarski, Dirk Rischke, Martin Savage, Thomas Sch\"afer, Achim Schwenk,
Igor Shovkovy, and Dam Son for discussions. I acknowledge the financial support of the German Academic Exchange Service (DAAD) and the hospitality of the Nuclear Theory Group at the University of Washington.


\end{document}